\documentclass[preprints,article,accept,moreauthors,pdftex,10pt,a4paper]{Definitions/mdpi}

\newcommand{\delete}[1]{}
\newcommand{\subX}{{\mbox{\tiny $X$}}}
\newcommand{\Dom}{\mathbb{X}}
\newcommand{\POVM}{\textsc{povm}}
\newcommand{\POVMs}{\textsc{povm}s}

\firstpage{1} 
\pubvolume{xx}
\issuenum{1}
\articlenumber{1}
\pubyear{2018}
\copyrightyear{2018}
\history{}

\pdfoutput=1

\usepackage{algorithm}
\usepackage{algorithmic}

\newcommand{\isdef}{\smash{\,\stackrel{\textrm{\scriptsize def}}{=}\,}}
\newcommand{\isdefdisp}{\stackrel{\textrm{def}}{=}}

\newtheoremstyle{thmstyle1}{3pt}{3pt}{}{}{}{.}{0.5em}{}
\newtheoremstyle{thmstyle2}{3pt}{3pt}{}{}{\bfseries}{.}{0.5em}{}
\theoremstyle{thmstyle2} \newtheorem{fact}{Fact}
\theoremstyle{thmstyle2}

\Title{Remote Sampling with Applications to General Entanglement Simulation}

\Author{Gilles Brassard $^{1,\dagger}$, Luc Devroye $^{2,\dagger}$ and Claude Gravel $^{3,\dagger}$}

\AuthorNames{Gilles Brassard, Luc Devroye and Claude Gravel}

\address{$^{1}$ \quad D\'epartement d'informatique et de recherche op\'erationnelle, Universit\'e de Montr\'eal, Canada, and Canadian~Institute for Advanced Research, Toronto, Canada; brassard@iro.umontreal.ca\\
$^{2}$ \quad School of Computer Science, McGill University, Canada; lucdevroye@gmail.com\\
$^{3}$ \quad D\'epartement d'informatique et de recherche op\'erationnelle, Universit\'e de Montr\'eal, Canada; claudegravel1980@gmail.com}

\firstnote{These authors contributed equally to this work; the order is alphabetic.}

\abstract{We show how to sample exactly discrete probability distributions whose defining parameters are distributed among remote parties. For this purpose, von Neumann's rejection algorithm is turned into a distributed sampling communication protocol. We~study the expected number of bits communicated among the parties and also exhibit a trade-off between the number of rounds of the rejection algorithm and the number of bits transmitted in the initial phase. Finally, we apply remote sampling to the simulation of quantum entanglement in its most general form possible, when an arbitrary number of parties share systems of arbitrary dimensions on which they apply arbi\-trary measurements (not~restricted to being projective measurements). In~case the dimension of the systems and the number of possible outcomes per party is bounded by a constant, it suffices to communicate an expected $O(m^2)$ bits in order to simulate exactly the outcomes that these measurements would have produced on those systems, where $m$ is the number of participants.}

\keyword{Communication complexity, quantum theory, classical simulation of entanglement, exact~sampling, random bit model, entropy.}

\begin{document}

\section{Introduction}\label{sc:introduction}

\begin{figure}[h!]
\begin{centering}
\includegraphics[trim = 17mm 190mm 20mm 33mm]{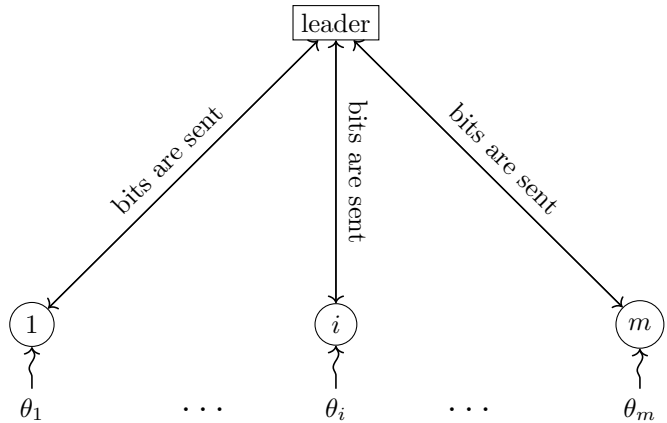}
\caption{The communication scheme}\label{fig:comscheme}\label{Fig:Comm:Scheme}
\end{centering}
\end{figure}

Let $\Dom$ be a nonempty finite set containing $n$ elements and \mbox{$p=(p_x)_{x\in\Dom}$} be a probability vector parameterized by some vector \mbox{$\theta=(\theta_1,\ldots,\theta_m)\in \Theta^{m}$} for an integer \mbox{$m \ge 2$}. For instance, the set $\Theta$ can be the real interval $[0,1]$ or the set of Hermitian semi-definite positive matrices as it is the case for the simulation of entanglement.
The probability vector $p$ defines a random variable $X$ such that $\mathbf{P}\{X=x\}\isdef p_x$ \mbox{for~$x\in\Dom$}\@.
To~sample exactly the probability vector $p$ means to produce an output $x$ such that $\mathbf{P}\{X=x\}=p_x$.
The problem of sampling probability distributions has been studied and is still studied extensively within different random and computational models. Here, we are interested in sampling \emph{exactly} a discrete distribution whose defining parameters are distributed among $m$ different parties.
The~$\theta_i$'s for $i\in\{1,\ldots,m\}$ are stored in $m$ different locations where the $i^{\text{th}}$ party holds~$\theta_i$. In~general, any communication topology between the parties would be allowed but in this work we concentrate for simplicity on a model in which we add a designated party known as the \emph{leader}, whereas the $m$ other parties are known as the \emph{custodians} because each of them is sole keeper of the corresponding parameter~$\theta$---hence there are \mbox{$m+1$} parties in total.
The~leader communicates in both directions with the custodians, who do not communicate among themselves. Allowing inter-custodian communication would not improve the communication efficiency of our scheme and can, at best, halve the number of bits communicated in any protocol. However, it could dramatically improve the sampling \emph{time} in a realistic model in which each party is limited to sending and receiving a fixed number of bits at any given time step, as demonstrated in our previous work~\cite{BrDeGr2015} concerning a special case of the problem considered here.
Graphically, the communication scheme is illustrated in Figure~\ref{Fig:Comm:Scheme}.

It may seem paradoxical that the leader can sample \emph{exactly} the probability vector $p$ with a \emph{finite} expected number of bits sent by the custodians, who may hold \emph{continuous} parameters that define~$p$. However, this counterintuitive possibility has been known to be achievable for more than a quarter-century in earlier work on the simulation of quantum entanglement by classical communication, starting with Refs.~\cite{maudlin92, bct99, steiner00} and culminating with our own Ref.~\cite{BrDeGr2015}.

Our protocol to sample remotely a given probability vector is presented in Section~\ref{sect:modifVN}. For~this purpose, the von Neumann rejection algorithm~\cite{vN51} is modified to produce an output \mbox{$x\in\Dom$} with exact probability~$p_x$ using mere approximations of those probabilities, which are computed based on partial knowledge of the parameters transmitted on demand by the custodians to the leader.
For~the sake of simplicity, and to concentrate on the new techniques, we assume initially that algebraic operations on real numbers can be carried out with infinite precision and that continuous random variables can be sampled.
Later, in Section~\ref{app:bitmodel}, we build on techniques developed in Ref.~\cite{BrDeGr2015} to obtain exact sampling in a realistic scenario in which
all computations are performed with finite precision.

In the intervening Section~\ref{sect:simquantent}, we study our motivating application of remote sampling, which is the simulation of quantum entanglement using classical resources and classical communication.\,\footnote{Readers who may not be interested in quantum information can still benefit from Section~\ref{sect:modifVN} and most of Section~\ref{app:bitmodel}, which make no reference to quantum theory in order to explain our general remote sampling strategies.}
A~special case of remote sampling has been used by the authors~\cite{BrDeGr2015}, in which the aim was to sample a specific probability distribution appearing often in quantum information science, namely the \mbox{$m$-par}\-tite GHZ distribution. More generally, consider a quantum system of dimension $d=d_1\cdots d_m$ represented by a density matrix $\rho$ known by the leader (surprisingly, the custodians have no need to know~$\rho$). Suppose that there are $m$ generalized measurements (\POVMs) acting on quantum systems of dimensions $d_1,\ldots, d_m$ whose possible outcomes lie in sets $\Dom_1, \ldots, \Dom_m$ of cardinality $n_1, \ldots, n_m$, respectively. Each custodian knows one and only one of the \POVMs{} and nothing else about the others.
The leader does not know initially any information about any of the \POVMs. Suppose in addition that the leader can generate independent identically distributed uniform random variables on the real interval~\mbox{$[0,1]$}.\,\footnote{\,In~Section~\ref{app:bitmodel}, we consider the more realistic scenario in which the only source of randomness comes from independent identically distributed uniform random \emph{bits}.}
We~show how to generate a random vector \mbox{$X= (X_1,\ldots, X_m)\in \Dom =\Dom_1\times \ldots \times \Dom_m$}
sampled from the exact joint probability distribution that would be obtained if each custodian $i$ had the $i^{\text{th}}$ share of $\rho$ (of~dimension~$d_i$) and measured it according to the $i^{\text{th}}$~\POVM, producing outcome \mbox{$x_i \in \Dom_i$}.
This task is defined formally in Section~\ref{sect:simquantent}, where we prove that the total expected number of bits transmitted between the leader and the custodians using remote sampling is $O(m^2)$ provided all the $d_i$'s and $n_i$'s are bounded by some constant.
The exact formula, involving $m$ as well as the $d_i$'s and $n_i$'s,
is given as Eq.~(\ref{eq:final_simul_tradeoff}) in Section~\ref{sect:simquantent}.
This result subsumes that of Ref.~\cite{BrDeGr2015} since
all $d_i$'s and $n_i$'s are equal to~2 for projective measurements on individual qubits of the \mbox{$m$-par}\-tite GHZ state.

\section{Remote Sampling}\label{sect:modifVN}

As explained in the Introduction, we show how to sample \emph{remotely} a discrete probability vector \mbox{$p=(p_x)_{x\in\Dom}$}. The task of sampling is carried by a \emph{leader} ignorant of some parameters \mbox{$\theta=(\theta_1,\ldots,\theta_m)$} that come in the definition of the probability vector, where each $\theta_i$ is known by the $i^{\text{th}}$ \emph{custodian} only, with whom the leader can communicate. We~strive to minimize the amount of communication required to achieve this task.

To solve our conundrum,
we modify the von Neumann rejection algorithm~\cite{vN51,Devroye1986}. Before explaining those modifications, let us review the original algorithm. Let \mbox{$q=(q_x)_{x\in\Dom}$} be a probability vector that we know how to sample on the same set $\Dom$, and let \mbox{$C\geq 1$} be such that \mbox{$p_x\leq C q_x$} for all \mbox{$x\in\Dom$}. The classical von Neumann rejection algorithm is shown as Algorithm~\ref{algo:vonNeumannOriginal}. It~is well-known that the expected number of times round the \textbf{repeat} loop is exactly~$C$.

\begin{algorithm}
\caption{Original von Neumann rejection algorithm}\label{algo:vonNeumannOriginal}
{\fontsize{10}{18}\selectfont
\begin{algorithmic}[1]
\LOOP
\STATE Sample $X$ according to $(q_x)_{x\in\Dom}$\label{line:X:original}
\STATE Sample $U$ uniformly on $[0,1]$\label{line:U:original}
\IF{ $UCq_{\subX} \leq p_{\subX}$}\label{line:vnorigtest}
\RETURN $X$ \COMMENT{$X$ is accepted}
\ENDIF
\ENDLOOP
\end{algorithmic}}
\end{algorithm}

If only partial knowledge about the parameters defining $p$ is known, it would seem that the condition in line~\ref{line:vnorigtest} cannot be decided. Nevertheless, the strategy is to build a sequence of increasingly accurate approximations that converge to the left and right sides of the test. As explained below, the number of bits transmitted depends on the number of bits needed to compute $q$, and on the accuracy in~$p$ required to accept or reject. This task can be achieved either in the \emph{random bit model}, in which only i.i.d.\ random bits are generated, or in the less realistic \emph{uniform model}, in which uniform continuous random variables are needed.
The random bit model was originally suggested by von Neumann~\cite{vN51}, but only later given this name and formalized by Knuth and Yao~\cite{KnuthYao1976}.
In~this section, we concentrate for simplicity on the uniform model, leaving the more practical random bit model for Section~\ref{app:bitmodel}.

\begin{Definition}
A \mbox{$t$-bit} \emph{approximation} of a real number $x$ is any $\hat{x}$ such that \mbox{$|x-\hat{x}|\leq 2^{-t}$}. A~special case of \mbox{$t$-bit} approximation is the \mbox{$t$-bit} \emph{truncation} \mbox{$\hat{x} = \mathrm{sign}(x) \lfloor |x| 2^{t} \rfloor / 2^{t}$}, where $\mathrm{sign}(x)$ is equal to $+1$, $0$ or $-1$ depending on the sign of~$x$.
If~$\alpha=a+bi$ is a complex number, where \mbox{$i=\sqrt{-1}$}, then a \mbox{$t$-bit} approximation (resp.~truncation) $\hat{\alpha}$ of $\alpha$ is any \mbox{$\hat{a}+\hat{b}i$}, where $\hat{a}$ and $\hat{b}$ are \mbox{$t$-bit} approximations (resp.~truncations) of $a$ and $b$, respectively.
\end{Definition}

Note that we assume without loss of generality that approximations of probabilities are always constrained to be real numbers between~$0$ and~$1$, which can be enforced by snapping any out-of-bound approximation (even if it's a complex number) to the closest valid value.

Consider an integer \mbox{$t_0>0$} to be determined later.
Our strategy is for the leader to compute
the probability vector \mbox{$q=(q_x)_{x\in\Dom}$} defined below, based on $t_0$-bit approximations $p_x(t_0)$ of the probabilities $p_x$ for each \mbox{$x\in\Dom$}.
For~this purpose, the leader receives sufficient information from the custodians to build the entire vector $q$ at the outset of the protocol.
This makes $q$ the ``easy-to-sample'' distribution required in von Neumann's technique,
which is easy not from a computational viewpoint, but in the sense that no further communication is required for the leader to sample it as many times as needed.
Let
\begin{equation}\label{equa:C}
C = \sum_{x}{\left(p_{x}(t_0)+2^{-t_0}\right)}
\end{equation}
and
\begin{equation}\label{equa:qiko}
q_x =  \left(p_{x}(t_0)+2^{-t_0}\right) / C \, .
\end{equation}
Noting that \mbox{$\sum_{x} q_{x} = 1$}, these $q_x$ define a proper
probability vector~\mbox{$q=(q_x)_{x\in\Dom}$}.
Using the definition of a $t$-bit approximation and the definition of $q_x$ from Eq.~(\ref{equa:qiko}), we have that
\begin{displaymath}
p_{x} \leq \Big( p_x(t_0)+2^{-t_0} = Cq_x \Big) \leq p_x + 2 \times 2^{-t_{0}} \, .
\end{displaymath}
Taking the sum over the possible values for $x$ and recalling that set $\Dom$ is of cardinality~$n$,
\begin{equation}
1\leq C\leq 1+2^{1-t_0} n \, .\label{equa:uppbndC}
\end{equation}

Consider any~\mbox{$x\in\Dom$} sampled according to~$q$ and $U$ sampled uniformly in~\mbox{$[0,1]$} as in lines~\ref{line:X:original} and~\ref{line:U:original} of  Algorithm~\ref{algo:vonNeumannOriginal}.
Should~$x$ be accepted because \mbox{$UCq_x \leq p_x$}, this can be certified by any \mbox{$t$-bit} approximation $p_x(t)$ of~$p_x$ such that
\mbox{$UCq_x \leq p_x(t)-2^{-t}$} for some positive integer~$t$ since \mbox{$p_x(t) \leq p_x+2^{-t}$}.
Conversely, any integer $t$ such that \mbox{$UCq_x > p_{x}(t)+2^{-t}$} certifies
that $x$ should be rejected because it implies that \mbox{$UCq_x > p_x$} since \mbox{$p_{x}(t) \geq p_{x}-2^{-t}$}.
On~the other hand, no~decision can be made concerning $UCq_x$ versus $p_x$ if \mbox{$- 2^{-t} < UCq_{x} - p_{x}(t) \leq 2^{-t}$}.
It~follows that one can modify Algorithm~\ref{algo:vonNeumannOriginal} above into Algorithm~\ref{algo:vnadaptedone} below,
in which a sufficiently precise approximation of $p_{x}$ suffices to make the correct
decision to accept or reject an $x$ sampled according to distribution~$q$.
A~well-chosen value of $t_0$ must be input into this algorithm, as discussed later.

\begin{algorithm}
\caption{Modified rejection algorithm -- Protocol for the leader}\label{algo:vnadaptedone}
{\fontsize{10}{18}\selectfont
\begin{algorithmic}[1]
\REQUIRE Value of $t_0$
\STATE Compute $p_x(t_0)$ for each $x \in \Dom$ \\[-1ex]
\COMMENT{The leader needs information from the custodians in order to compute these approximations}
\STATE Compute $C$ and $q=(q_x)_{x\in\Dom}$ as per Eqs.~(\ref{equa:C}) and~(\ref{equa:qiko})
\STATE Sample $X$ according to $q$\label{gen:X}
\STATE Sample $U$ uniformly on $[0,1]$\label{gen:unif}
\FOR{$t=t_0$ \TO $\infty$\label{for:loop}}
\IF{ $UCq_{\subX} \leq p_{\subX}(t)-2^{-t}$}\label{test:to:accept:continuous}
\RETURN $X$\label{line:accept} \COMMENT{$X$ is accepted}
\ELSIF{$UCq_{\subX} > p_{\subX}(t)+2^{-t}$}\label{test:to:reject:continuous}
\STATE \textbf{go to}~\ref{gen:X} \COMMENT{$X$ is rejected}
\ELSE
\STATE Continue the \textbf{for} loop\label{line:no:decision} \\[-1ex]
\COMMENT{We cannot decide whether to accept or reject because $- 2^{-t} < UCq_{x} - p_{x}(t) \leq 2^{-t}$\,; \\[-1ex]
communication may be required in order for the leader to compute \mbox{$p_{\subX}(t+1)$}\,; \\[-1ex]
it~could be that bits previously communicated to compute $p_{x}(t)$ can be reused.}\label{we:can:reuse}
\ENDIF
\ENDFOR
\end{algorithmic}}
\end{algorithm}

\begin{Theorem}\label{thm:main}
Algorithm~\ref{algo:vnadaptedone} is correct, i.e., it terminates and returns $X=x$ with probability $p_x$. Furthermore, let $T$ be the random variable that denotes the value of variable $t$ upon termination of any instance of the \textup{\textbf{for}} loop, whether the loop terminates in rejection or acceptation. Then
\begin{equation}\label{eq:thm1}
\mathbf{E}(T)\leq t_0 + 3 \, .
\end{equation}
\end{Theorem}

\begin{proof}
Consider any \mbox{$x\in\Dom$} and \mbox{$t \geq t_0$}.
To~reach \mbox{$T>t$}, it must be that
\mbox{$- 2^{-t} < UCq_{x} - p_{x}(t) \leq 2^{-t}$}.
Noting that \mbox{$q_x \neq 0$} according to Eq.~(\ref{equa:qiko}),
the probability that \mbox{$T>t$} when \mbox{$X=x$} is therefore upper-bounded as follows:
\begin{align}
\mathbf{P}\!\left\{T>t~\rvert~X=x\right\}
& \leq \mathbf{P}\!\left\{- 2^{-t} < UCq_{x} - p_{x}(t) \leq 2^{-t} \right\} \nonumber \\
&= \mathbf{P}\!\left\{ \frac{p_{x}(t) - 2^{-t}}{Cq_x} <  U \leq \frac{p_{x}(t) + 2^{-t}}{Cq_x} \right\} \nonumber \\
&\le \frac{p_{x}(t) + 2^{-t}}{Cq_x} - \frac{p_{x}(t) - 2^{-t}}{Cq_x}
= \frac{2 \times 2^{-t}}{Cq_x} \leq 2^{t_0-t+1} \, . \label{prob_nohalt}
\end{align}
The last inequality uses the fact that
\begin{displaymath}
Cq_x = p_x(t_0)+2^{-t_0} \geq 2^{-t_0} \, .
\end{displaymath}

It follows that the probability that more turns round the \textbf{for} loop are required decreases exponentially with each new turn once \mbox{$t > t_0+1$},
which suffices to guarantee termination of the \textbf{for} loop with probability~$1$.
Termination of the algorithm itself comes from the fact that the choice of $X$ and $U$ in lines~\ref{gen:X} and~\ref{gen:unif} leads to acceptance at line~\ref{line:accept}---and therefore termination---with probability $1/C$, as demonstrated by von Neumann in the analysis of his rejection algorithm.

The fact that \mbox{$X=x$} is returned with probability $p_x$ is an immediate consequence of the correctness of the von Neumann rejection algorithm since our adaptation of this method to handle the fact that only approximations of~$p_{\subX}$ are available does not change the decision to accept or reject any given candidate sampled according to~$q$.

In order to bound the expectation of $T$, we note that
\mbox{$\mathbf{P}\{T>t~\rvert~X=x\} = 1$} when \mbox{$t < t_0$}
since we start the \textbf{for} loop at \mbox{$t = t_0$}.
We~can also use vacuous \mbox{$\mathbf{P}\{T>t_0~\rvert~X=x\} \le 1$} rather
than the worse-than-vacuous upper bound of $2$ given by Eq.~(\ref{prob_nohalt}) in the case \mbox{$t=t_0$}.
Therefore
\begin{align*}
\mathbf{E}(T~\rvert~X=x)&=\sum_{t=0}^{\infty}{\mathbf{P}\{T>t~\rvert~ X=x\}}\\
&=\sum_{t=0}^{t_0}{\mathbf{P}\{T>t~\rvert~ X=x\}} + \!\!\!\!\sum_{t=t_0+1}^{\infty}\!\!{\mathbf{P}\{T>t~\rvert~ X=x\}}\\
&\leq t_0 + 1 + 2^{t_0+1}\!\!\!\!\sum_{t=t_{0}+1}^{\infty}\!\!{2^{-t}} = t_0+3 \, .
\end{align*}
It remains to note that since \mbox{$\mathbf{E}(T~\rvert~X=x) \leq t_0+3$} for all \mbox{$x\in\Dom$}, it follows that \mbox{$\mathbf{E}(T) \leq t_0+3$} without condition.
\end{proof}

Let $S$ be the random variable that represents the number of times variable $X$ is sampled according to $q$ at line~\ref{gen:X}, and let $T_i$ be the random variable that represents the value of variable $T$ upon termination of the $i^{\text{th}}$ instance of the \textbf{for} loop starting at line~\ref{for:loop}, for $i\in\{1,\ldots,S\}$.
The random variables $T_i$ are independently and identically distributed as the random variable $T$ in Theorem~\ref{thm:main} and the expected value of $S$ is~$C$.
Let~$X_{1},\ldots, X_{S}$ be the random variables chosen at successive passes at line~\ref{gen:X}, so~that $X_1,\ldots, X_{S-1}$ are rejected, whereas $X_{S}$ is returned as the final result of the algorithm.

To analyse the communication complexity of Algorithm~\ref{algo:vnadaptedone}, we introduce function $\gamma_{x}(t)$ for each \mbox{$x\in\Dom$} and \mbox{$t>t_0$}, which denotes the \emph{incremental} number of bits that
the leader must receive from the custodians in order to compute $p_x(t)$, taking account of the information that may already be available if he had previously computed \mbox{$p_x(t-1)$}.
For completeness, we include in~$\gamma_{x}(t)$ the cost of the communication required for the leader to request more information from the custodians.
We~also introduce function $\delta(t)$ for $t \geq 0$, which denotes the number of bits that
the leader must receive from the custodians in order to compute $p_x(t)$ for all~\mbox{$x\in\Dom$} in a ``simultaneous'' manner. Note that it could be much less expensive to compute those $n$ values than $n$ times the cost of computing any single one of them because some of the parameters held by the custodians may be relevant to more than one of the~$p_x$'s. 
The total number of bits communicated in order to implement Algorithm~\ref{algo:vnadaptedone} is therefore given by random variable
\begin{displaymath}
Z = \delta(t_0)+\sum_{i=1}^{S} \, \sum_{t=t_0+1}^{T_i} {\gamma_{\subX_{i}}(t)} \, .
\end{displaymath}
For simplicity, let us define function $\gamma(t) \isdef \max_{x\in\Dom}\gamma_x(t)$.
We~then have
\begin{displaymath}
Z \leq \delta(t_0)+\sum_{i=1}^{S} \, \sum_{t=t_0+1}^{T_i} {\gamma(t)} \, ,
\end{displaymath}
whose expectation, according to Wald's identity, is
\begin{equation}\label{eq:if:no:trunc}
\mathbf{E}(Z) \leq \delta(t_0) + \mathbf{E}(S) \, \mathbf{E}
\Bigg( \sum_{t=t_0+1}^{T} {\gamma(t)} \Bigg) \, .
\end{equation}
Assuming the value of $\gamma(t)$ is upper-bounded by some~$\gamma$,
\begin{align}
\mathbf{E}(Z) & \leq \delta(t_0) + \mathbf{E}(S) \mathbf{E}(T-t_0) \gamma \nonumber \\
& \leq \delta(t_0) + 3 \gamma C \nonumber \\
& \leq \delta(t_0) + 3 \gamma \big( 1+2^{1-t_0} n \big)  \label{eq:tradeoff}
\end{align}
because \mbox{$\mathbf{E}(S) = C$} and using Eqs.~(\ref{eq:thm1}) and~(\ref{equa:uppbndC}).

Depending on the specific application, which determines $\gamma$ and function~$\delta(t)$, Eq.~(\ref{eq:tradeoff}) is key to a trade-off that can lead to an optimal choice of~$t_0$ since a larger $t_0$ decreases  $2^{1-t_0}$ but is likely to increase~$\delta(t_0)$. The value of $\gamma$ may play a r\^ole in the balance. The next section, in which we consider the simulation of quantum entanglement by classical communication, gives an example of this trade-off in action.

\section{Simulation of Quantum Entanglement Based on Remote Sampling}\label{sect:simquantent}

Before introducing the simulation of entanglement, let us establish some notation and mention the mathematical objects that we shall need. It is assumed that the reader is familiar with linear algebra, in particular the notion of a semi-definite positive matrix, Hermitian matrix, trace of a matrix, tensor product, etc. For a discussion about the probabilistic and statistical nature of quantum theory, see Ref.~\cite{Holevo2001}.
For~convenience, we use $[n]$ to denote the set \mbox{$\{1,2,\ldots,n\}$} for any integer~$n$.

Consider integers $m, d_1, d_2, \ldots, d_m, n_1, n_2, \ldots, n_m$, all greater than or equal to~$2$.
Define \mbox{$d=\prod_{i=1}^{m}{d_i}$} and \mbox{$n=\prod_{i=1}^{m}{n_i}$}.
Let $\rho$ be a \mbox{$d\times d$} density matrix.
Recall that any density matrix is Hermitian, semi-definite positive and unit-trace, which implies that its diagonal elements are real numbers between $0$ and~$1$.
For~each $i \in [m]$ and $j \in [n_i]$, let $M_{ij}$ be a \mbox{$d_i \times d_i$} Hermitian semi-definite positive matrix such that
\begin{equation}\label{eq:povm}
\sum_{j\in[n_i]}{M_{ij}}=I_{d_i} \, ,
\end{equation}
where $I_{d_{i}}$ is the \mbox{$d_{i}\times d_{i}$} identity matrix. In other words, each set $\{M_{ij}\}_{j\in[n_i]}$ is a \POVM{} (positive-operator valued measure)~\cite{Holevo2001}.

As introduced in Section~\ref{sc:introduction}, we consider one \emph{leader} and $m$ \emph{custodians}.
They all know density matrix~$\rho$, but only the $i^{\text{th}}$ custodian knows the $i^{\text{th}}$ \POVM, meaning that he knows the matrices $M_{ij}$ for all
\mbox{$j \in [n_i]$}.
If~a physical system of dimension $d$ in state $\rho$ were shared between the custodians, in the sense that the $i^{\text{th}}$ custodian had possession of the $i^{\text{th}}$ subsystem of dimension~$d_i$, each custodian could perform locally his assigned \POVM{} and output the outcome, an integer between~$1$ and~$n_i$. The~joint output would belong to \mbox{$\Dom \;\isdef\; [n_1] \times [n_2] \times \cdots \times [n_m]$}, a set of cardinality~$n$, sampled according to the probability distribution stipulated by the laws of quantum theory, which we review below.

Our~task is to sample $\Dom$ with the exact same probability distribution even though there is no physical system in state $\rho$ available to the custodians, and in fact all parties considered are purely classical! We~know from Bell's Theorem~\cite{Bell64} that this task is impossible in general without communication, even when \mbox{$m=2$}, and our goal is to minimize the amount of communication required to achieve~it.
Special cases of this problem have been studied extensively for expected~\cite[etc\@.]{maudlin92, steiner00} and worst-case~\cite[etc\@.]{bct99,TB03} communication complexity,
but here we solve it in its most general setting, albeit only in the expected sense.
For~this purpose, the leader will centralize the operations while requesting as little information as possible from the custodians on their assigned \POVMs.
Once the leader has successfully sampled \mbox{$X= (X_1,\ldots, X_m)$}, he transmits each $X_i$ to the $i^{\text{th}}$ custodian, who can then output it as would have been the case had quantum measurements actually taken place.

We now review the probability distribution $\Dom$ that we need to sample, according to quantum theory.
For each vector $x=(x_1,\ldots,x_{m}) \in \Dom$,
let $M_{x}$ be the \mbox{$d \times d$} tensor product of matrices $M_{ix_{i}}$ for each $i \in [m]$:
\begin{equation}
M_{x} = \bigotimes_{i=1}^{m}{M_{ix_{i}}} \, .\label{eqn:Mx}
\end{equation}
The set $\{M_x\}_{x\in\Dom}$ forms a global \POVM{} of dimension $d$, which applied to density matrix $\rho$ defines a joint probability vector on~$\Dom$.
The probability $p_x$ of obtaining any \mbox{$x=(x_1,\ldots,x_m)\in\Dom$} is given by
\begin{equation}
p_{x}=\text{Tr}\big(\rho M_{x}\big)=\text{Tr}\Bigg(\rho\Bigg(\bigotimes_{i=1}^{m}{M_{ix_{i}}}\Bigg)\Bigg) \, .\label{eqn:probx}
\end{equation}

For a matrix $A$ of size $s \times s$ and any pair of indices $r$ and $c$ between $0$ and~\mbox{$s-1$}, we use $(A)_{rc}$ to denote the entry of $A$ located in the $r^{\text{th}}$ row and $c^{\text{th}}$ column.
Matrix indices start at~$0$ rather than~$1$ to facilitate Fact~\ref{fact:correspondance_rc} below.
We~now state various facts for which we provide cursory justifications since they follow from elementary linear algebra and quantum theory, or they are lifted from previous work.

\begin{fact}
For all $x \in \Dom$, we have \mbox{$0 \leq p_x \leq 1$} when $p_x$ is defined according to Eq.~(\ref{eqn:probx}); furthermore, \mbox{$\sum_{x\in\Dom}{p_{x}}=1$}.
This is obvious because
quantum theory tells us that Eq.~(\ref{eqn:probx}) defines the probability distribution over all possible outcomes \mbox{$x\in\Dom$} of the joint measurement. Naturally, this statement could also be proven from Eqs.~(\ref{eq:povm}) and~(\ref{eqn:probx}) using elementary linear algebra.
\end{fact}

\begin{fact}\label{fact:correspondance_rc}
For each $x=(x_1,\ldots,x_m) \in \Dom$, matrix $M_{x}$ is the tensor product of $m$ matrices as given in Eq.~(\ref{eqn:Mx}). Therefore, each entry $(M_{x})_{rc}$ is the product of $m$ entries of the $M_{ix_{i}}$'s. Specifically, consider any indices $r$ and $c$ between $0$ and~\mbox{$d-1$} and let $r_i$ and $c_i$ be the indices between $0$ and~\mbox{$d_{i}-1$}, for each \mbox{$i\in[m]$}, such that
\begin{align*}
r&=r_{1}+r_{2}d_{1}+r_{3}d_{1}d_{2}+\ldots+r_{m}d_{1}\cdots d_{m-1} \\
c&=c_{1}+c_{2}d_{1}+c_{3}d_{1}d_{2}+\ldots+c_{m}d_{1}\cdots d_{m-1} \, .
\end{align*}
The $r_i$'s and $c_i$'s are uniquely defined by the principle of mixed-radix numeration. We~have
\begin{displaymath}
(M_{x})_{rc}=\prod_{i=1}^{m}{\big(M_{ix_{i}}\big)_{r_{i}c_{i}}} \, .
\end{displaymath}
\end{fact}

\begin{fact}\label{fact:sylvester}
Let $M$ be a Hermitian semi-definite positive matrix.
Every entry $(M)_{ij}$ of the matrix satisfies
\begin{displaymath}
|(M)_{ij}|\leq \sqrt{(M)_{ii}(M)_{jj}} \, .
\end{displaymath}
This follows from the fact that all principal submatrices of any Hermitian semi-definite positive matrix are semi-definite positive~\cite[Observation~7.1.2, page 430]{horn2012matrix}. In~particular the principal submatrix
\begin{displaymath}
\left( \begin{matrix} (M)_{ii} & (M)_{ij} \\ (M)_{ji} & (M)_{jj} \end{matrix} \right)
\end{displaymath}
is semi-definite positive, and therefore it has nonnegative determinant:
\begin{displaymath}
(M)_{ii} (M)_{jj} - (M)_{ij} (M)_{ji} = (M)_{ii} (M)_{jj} - (M)_{ij} (M)_{ij}^{*} = (M)_{ii} (M)_{jj} - | (M)_{ij} |^2 \geq 0
\end{displaymath}
by virtue of $M$ being Hermitian, where $\alpha^{*}$ denotes the complex conjugate of~$\alpha$.
\end{fact}

\begin{fact}\label{fact:boundedmodulus}
The norm $|(\rho)_{ij}|$ of any entry of a density matrix $\rho$ is less than or equal to~$1$.
This follows directly from Fact~\ref{fact:sylvester} since density matrices are Hermitian semi-definite positive, and from the fact that diagonal entries of density matrices, such as $(\rho)_{ii}$ and~$(\rho)_{jj}$, are real values between $0$ and~$1$.
\end{fact}

\begin{fact}\label{fact:boundedentrypovm}
Given any \POVM{} $\{M_{\ell}\}_{\ell=1}^{L}$, we have that
\begin{enumerate}
\item \mbox{$0 \leq (M_{\ell})_{ii}\leq 1$} for all $\ell$ and~$i$, and
\item \mbox{$|(M_{\ell})_{ij}| \leq 1$} for all $\ell$, $i$ and~$j$.
\end{enumerate}
The~first statement follows from the fact that \mbox{$\sum_{\ell=1}^{L} {M_{\ell}}$} is the identity matrix
by definition of \POVMs,
and therefore \mbox{$\sum_{\ell=1}^{L} ({M_{\ell}})_{ii} = 1$} for all~$i$,
and the fact that each \mbox{$({M_{\ell}})_{ii} \geq 0$} because each $M_{\ell}$ is semi-definite positive.
The~second statement follows from the first by applying Fact~\ref{fact:sylvester}.
\end{fact}

\begin{fact}[This is a special case of Theorem~1 from Ref.~\protect{\cite{BrDeGr2015}}, with~\mbox{$v=0$}]\label{fact:IEEEthm}
Let \mbox{$k \ge 1$} be an integer and consider any two real numbers $a$ and~$b$.
If~$\hat{a}$ and~$\hat{b}$ are arbitrary \mbox{$k$-bit} approximations
of $a$ and $b$, respectively, then \mbox{$\hat{a}+\hat{b}$} is a \mbox{$(k-1)$-bit} approximation of \mbox{$a+b$}.
If,~in addition, $a$ and~$b$ are known to lie in interval \mbox{$[-1,1]$},
which can also be assumed without loss of generality concerning $\hat{a}$ and $\hat{b}$ since otherwise they can be safely pushed back to the appropriate frontier of this interval,
then $\hat{a}\hat{b}$ is a \mbox{$(k-1)$-bit} approximation of~$ab$.
\end{fact}

\begin{fact}\label{fact:approx_complex}
Let \mbox{$k \ge 1$} be an integer and consider any two \emph{complex} numbers $\alpha$ and $\beta$.
If~$\hat{\alpha}$ and~$\hat{\beta}$ are arbitrary \mbox{$k$-bit} approximations of $\alpha$ and $\beta$, respectively, then \mbox{$\hat{\alpha}+\hat{\beta}$} is a \mbox{$(k-1)$-bit} approximation of \mbox{$\alpha+\beta$}.
If,~in addition, \mbox{$k \ge 2$} and 
the real and imaginary parts of $\alpha$ and $\beta$ are known to lie in interval \mbox{$[-1,1]$}, which can also be assumed without loss of generality concerning $\hat{\alpha}$ and $\hat{\beta}$, then $\hat{\alpha}\hat{\beta}$ is a \mbox{$(k-2)$-bit} approximation of $\alpha\beta$.
This is a direct consequence of Fact~\ref{fact:IEEEthm} in the case of addition.
In~the case of multiplication, consider \mbox{$\alpha=a+bi$}, \mbox{$\beta=c+di$},
\mbox{$\hat{\alpha}=\hat{a}+\hat{b}i$} and \mbox{$\hat{\beta}=\hat{c}+\hat{d}i$}, so that 
\begin{displaymath}
{\alpha}{\beta}=({a}{c}-{b}{d}) + ({a}{d}+{b}{c})i
\text{~~~and~~~}
\hat{\alpha}\hat{\beta}=(\hat{a}\hat{c}-\hat{b}\hat{d}) + (\hat{a}\hat{d}+\hat{b}\hat{c})i \, .
\end{displaymath}
By~the multiplicative part of Fact~\ref{fact:IEEEthm}, $\hat{a}\hat{c}$, $\hat{b}\hat{d}$, $\hat{a}\hat{d}$ and $\hat{b}\hat{c}$ are \mbox{$(k-1)$-bit} approximations of $ac$, $bd$, $ad$ and $bc$, respectively;
and then by the additive part of the same fact (which obviously applies equally well to subtraction), \mbox{$\hat{a}\hat{c}-\hat{b}\hat{d}$} and \mbox{$\hat{a}\hat{d}+\hat{b}\hat{c}$} are \mbox{$(k-2)$-bit} approximations of \mbox{$ac-bd$} and \mbox{$ad+bc$}, respectively.
\end{fact}

\begin{fact}[This is Corollary~2 from Ref.~\protect{\cite{BrDeGr2015}}]\label{fact:IEEEcor}
Let \mbox{$m \geq 2$} and \mbox{$k \geq \lceil \, \lg m \rceil$} be integers and let $\{a_j\}_{j=1}^m$ and $\{\hat{a}_j\}_{j=1}^m$ be real numbers
and their \mbox{$k$-bit} approximations, respectively, all in interval \mbox{$[-1,1]$}.
Then~\mbox{$\prod_{j=1}^m \hat{a}_j$ is a \mbox{$(k-\lceil \, \lg m \rceil)$-bit}} approximation of $\prod_{j=1}^m a_j$\,.
\end{fact}

\begin{fact}\label{fact:mult_complex}
Let \mbox{$m \geq 2$} and \mbox{$k \geq 2 \lceil \, \lg m \rceil$} be integers and let $\{\alpha_j\}_{j=1}^m$ and $\{\hat{\alpha}_j\}_{j=1}^m$
be complex numbers
and their \mbox{$k$-bit} approximations, respectively.
Provided it is known that \mbox{$|\alpha_j| \leq 1$} for each \mbox{$j\in[m]$},
a~\mbox{$(k- 2 \lceil \, \lg m \rceil)$-bit} approximation of $\prod_{j=1}^m \alpha_j$
can be computed from knowledge of the~$\hat{\alpha}_j$'s.
The~proof of this fact follows essentially the same template as Fact~\ref{fact:IEEEcor}, except that \emph{two} bits of precision may be lost at each level up the binary tree introduced in Ref.~\cite{BrDeGr2015}, due to the difference between Facts~\ref{fact:IEEEthm} and~\ref{fact:approx_complex}.
A~subtlety occurs in the need for Fact~\ref{fact:approx_complex} to apply that the real
and imaginary parts of all the complex numbers
under consideration must lie in interval \mbox{$[-1,1]$}. This is automatic for the exact values since the $\alpha_j$'s are upper-bounded in norm by~1 and the product of such-bounded complex numbers is also upper-bounded in norm by~1,
which implies that their real and imaginary parts lie in interval \mbox{$[-1,1]$}.
For the approximations, however, we cannot force their \emph{norm} to be bounded by~1
because we need the approximations to be rational for communication purposes.
Fortunately, we can force the real and imaginary parts of all approximations computed
at each level up the binary tree to lie in interval \mbox{$[-1,1]$} because we know that they approximate such-bounded numbers.
Note that the product of two complex numbers whose real and imaginary parts
lie in interval \mbox{$[-1,1]$}, such as \mbox{$1+2^{-k}i$} and \mbox{$1-2^{-k}i$}, may not have this property, even if they are \mbox{$k$-bit} approximations of numbers bounded in norm by~1.
\end{fact}

\begin{fact}\label{fact:add_complex}
Let \mbox{$s \geq 2$} and \mbox{$k \geq \lceil \, \lg s \rceil$} be integers and let $\{\alpha_j\}_{j=1}^s$ and $\{\hat{\alpha}_j\}_{j=1}^s$ be complex numbers
and their \mbox{$k$-bit} approximations, respectively, without any restriction on their norm.
Then $\sum_{j=1}^s \hat{\alpha}_j$ is a \mbox{$(k- \lceil \, \lg s \rceil)$-bit} approximation of $\sum_{j=1}^s \alpha_j$\,.
Again, this follows the same proof template as Fact~\ref{fact:IEEEcor}, substituting multiplication of real numbers by addition of complex numbers, which allows us to drop any condition on the size of the numbers considered.
\end{fact}

\begin{fact}\label{fact:init_precision}
Consider any \mbox{$x=(x_1,\ldots,x_m)\in\Dom$} and any positive integer~$t$.
In~order to compute a \mbox{$t$-bit} approximation of~$p_x$, it suffices to have
\mbox{$(t+1+\lceil2\lg d\rceil+2\lceil\,\lg m\rceil)$-bit} approximations of each entry of the $M_{ix_i}$ matrices for all \mbox{$i \in [m]$}.
This is because
\begin{align}
p_{x}&=\text{Tr}(\rho M_x)
=\sum_{r=0}^{d-1}{(\rho M_{x})_{rr}}\nonumber\\
&=\sum_{r=0}^{d-1}{\sum_{c=0}^{d-1}{(\rho)_{rc}(M_{x})_{cr}}}\nonumber\\
&=\sum_{r=0}^{d-1}{\sum_{c=0}^{d-1}{(\rho)_{rc}\prod_{i=1}^{m}{(M_{ix_{i}})_{c_{i}r_{i}}}}}\label{eqn:precision_probx}
\end{align}
by virtue of Fact~\ref{fact:correspondance_rc}.
Every term of the double sum in Eq.~(\ref{eqn:precision_probx}) involves a product of $m$ entries, one per \POVM{} element, and therefore incurs a loss of at most $2\lceil\,\lg m\rceil$ bits of precision by Fact~\ref{fact:mult_complex}, whose condition holds thanks to Fact~\ref{fact:boundedentrypovm}. An~additional bit of precision may be lost in the multiplication by $(\rho)_{rc}$, even though that value is available with arbitrary precision
(and is upper-bounded by~$1$ in norm by Fact~\ref{fact:boundedmodulus}) because of the additions involved in multiplying complex numbers.
Then, we have to add \mbox{$s=d^2$} terms, which incurs an additional loss of at most \mbox{$\lceil\,\lg s\rceil = \lceil 2 \lg d \rceil$} bits of precision by Fact~\ref{fact:add_complex}.
In~total, \mbox{$(t+1+\lceil2\lg d\rceil+2\lceil\,\lg m\rceil)$-bit} approximations of the $(M_{ix_{i}})_{c_{i}r_{i}}$'s will result in a \mbox{$t$-bit} approximation of~$p_x$.
\end{fact}

\begin{fact}\label{fact:total_precision}
The leader can to compute $p_{x}(t)$ for any specific $x=(x_1,\ldots,x_m)\in\Dom$ and integer $t$ if he receives a total of
\begin{displaymath}
(t+2+\lceil2\lg d\rceil+2 \lceil\lg m\rceil)\sum_{i=1}^{m}{d_{i}^{2}}
\end{displaymath}
bits from the custodians. This is because the $i^{\text{th}}$ custodian has the description of matrix $M_{ix_i}$ of size $d_i\times d_i$, which is defined by exactly $d_{i}^{2}$ \emph{real} numbers since the matrix is Hermitian.
By~virtue of Fact~\ref{fact:init_precision}, it is sufficient for the leader to have 
\mbox{$(t+1+\lceil2\lg d\rceil+2\lceil\,\lg m\rceil)$-bit} approximations for all those $\sum_{i=1}^{m}{d_{i}^{2}}$ numbers. Since each one of them lies in interval \mbox{[-1,1]} by Fact~\ref{fact:boundedentrypovm}, well-chosen \mbox{$k$-bit} approximations (for~instance \mbox{$k$-bit} truncations) can be conveyed by the transmission of \mbox{$k+1$} bits, one of which carries the sign.

Note that the \mbox{$t$-bit} approximation of $p_x$ computed according to Fact~\ref{fact:total_precision}, say \mbox{$a+bi$}, may very well have a nonzero imaginary part~$b$, albeit necessarily between $-2^{-t}$ and~$2^{-t}$. Since $p_x(t)$ must be a real number between $0$ and~$1$, the leader sets \mbox{$p_x(t) = \max(0,\min(1,a))$}, taking no account of~$b$, although a paranoid leader may wish to test that \mbox{$-2^{-t} \leq b \leq 2^{-t}$} indeed and raise an alarm in case it is~not (which of course is mathematically impossible unless the custodians are not given proper \POVMs, unless they misbehave, or unless a computation or communication error has occurred).
\end{fact}

\begin{fact}\label{fact:total_precision_for_all_x}
For any $t$, the leader can compute $p_{x}(t)$ for each and every $x\in\Dom$ if he receives
\begin{displaymath}
\delta(t) \isdefdisp (t+2+\lceil2\lg d\rceil+2 \lceil\lg m\rceil)\sum_{i=1}^{m}{n_i d_{i}^{2}}
\end{displaymath}
bits from the custodians. This is because it suffices for each custodian $i$ to send to the leader
\mbox{$(t+1+\lceil2\lg d\rceil+2\lceil\,\lg m\rceil)$-bit} approximations of all $n_i d_{i}^{2}$ real numbers that define the entire $i^{\text{th}}$ \POVM, i.e., all the matrices $M_{ij}$ for $j\in[n_i]$. This is a nice example of the fact that it may be much less expensive for the leader to compute at once $p_x(t)$ for all \mbox{$x\in\Dom$}, rather than computing them one by one independently, which would cost 
\begin{displaymath}
n (t+2+\lceil2\lg d\rceil+2 \lceil\lg m\rceil)\sum_{i=1}^{m}{d_{i}^{2}} =
(t+2+\lceil2\lg d\rceil+2 \lceil\lg m\rceil)\sum_{i=1}^{m}{n d_{i}^{2}} \gg \delta(t)
\end{displaymath}
bits of communication by applying $n$ times Fact~\ref{fact:total_precision}.
\end{fact}

After all these preliminaries, we are now ready to adapt the general template of Algorithm~\ref{algo:vnadaptedone} to our entanglement-simulation conundrum.
We~postpone the choice of~$t_0$ until the communication complexity analysis of the new algorithm.

\begin{algorithm}
\caption{Protocol for simulating arbitrary entanglement subjected to arbitrary measurements}\label{algo:simulation}
{\fontsize{10}{18}\selectfont
\begin{algorithmic}[1]
\STATE Each custodian $i\in[m]$ sends his value of $n_i$ to the leader, who computes $n=\prod_{i=1}^m n_i$\label{line_one}
\STATE The leader chooses $t_0$ and informs the custodians of its value\label{line_two}
\STATE Each custodian $i\in[m]$ sends to the leader $(t_0+1+\lceil2\lg d \rceil+2\lceil\lg m\rceil)$-bit truncations \\[-1ex]
of the real and imaginary parts of the entries defining matrix $M_{ij}$ for each $j\in[n_{i}]$\label{use:truncations}
\STATE The leader computes $p_x(t_0)$ for every $x\in\Dom$, using Fact~\ref{fact:total_precision_for_all_x}\label{comp:pxt0:ent}
\STATE The leader computes $C$ and $q=(q_x)_{x\in\Dom}$ as per Eqs.~(\ref{equa:C}) and~(\ref{equa:qiko})
\STATE $\textsf{accept} \leftarrow \textsf{false}$
\REPEAT
\STATE $\textsf{reject} \leftarrow \textsf{false}$
\STATE The leader samples $X=(X_1,X_2,\ldots,X_m)$ according to $q$\label{gen:X:ent}
\STATE The leader informs each custodian $i\in[m]$ of the value of~$X_i$\label{line:informs}
\STATE The leader samples $U$ uniformly on $[0,1]$\label{gen:unif:ent}
\STATE $t \leftarrow t_0$
\REPEAT
\IF{ $UCq_{\subX} \leq p_{\subX}(t)-2^{-t}$}
\STATE $\textsf{accept} \leftarrow \textsf{true}$ \COMMENT{$X$ is accepted}
\ELSIF{$UCq_{\subX}> p_{\subX}(t)+2^{-t}$}
\STATE $\textsf{reject} \leftarrow \textsf{true}$ \COMMENT{$X$ is rejected}
\ELSE
\STATE The leader asks each custodian $i\in[m]$ for one more bit in the truncation\\[-1ex]
of the real and imaginary parts of the entries defining matrix $M_{i\subX_i}$;\label{one:more:bit}
\STATE Using this information, the leader updates $p_{\subX}(t)$ into $p_{\subX}(t+1)$;\label{line:update}
\STATE $t \leftarrow t+1$
\ENDIF
\UNTIL \textsf{accept} \textbf{or} \textsf{reject}
\UNTIL \textsf{accept} 
\STATE The leader requests each custodian $i\in[m]$ to output his $X_i$\label{line:final}
\end{algorithmic}}
\end{algorithm}

\newpage  

To analyse the expected number of bits of communication required by this algorithm, we apply Eq.~(\ref{eq:tradeoff}) from Section~\ref{sect:modifVN} after defining explicitly the cost parameters $\delta(t_0)$ for the initial computation of $p_x(t_0)$ for all \mbox{$x\in\Dom$} at lines~\ref{use:truncations} and~\ref{comp:pxt0:ent}, and $\gamma$ for the upgrade from a specific $p_\subX(t)$ to \mbox{$p_\subX(t+1)$} at lines~\ref{one:more:bit} and~\ref{line:update}.
For~simplicity, we shall ignore the negligible amount of communication entailed at line~\ref{line_one} (which is \mbox{$\sum_{i=1}^{m}{\lceil\,\lg n_i \rceil} \leq m + \lg n$} bits), line~\ref{line_two} ($\lceil\,\lg t_0 \rceil$ bits), line~\ref{line:informs} (also \mbox{$\sum_{i=1}^{m}{\lceil\,\lg n_i \rceil}$} bits, but repeated \mbox{$\mathbf{E}(S) \leq 1+2^{1-t_0} n$} times) and line~\ref{line:final} (m~bits)
because they are not taken into account in Eq.~(\ref{eq:tradeoff}) since they are absent from Algorithm~\ref{algo:vnadaptedone}.
If~we counted it all, this would add $O((1+2^{1-t_0} n) \lg n + \lg t_0)$ bits to Eq.~(\ref{eq:final_simul}) below, which would be less than $10\lg n$ bits added to Eq.~(\ref{eq:final_simul_tradeoff}),
with no effect at all~on~Eq.~(\ref{eq:bounded}).

According to Fact~\ref{fact:total_precision_for_all_x},
\begin{displaymath}
\delta(t_0)=(t_0+2+\lceil2\lg d\rceil+2 \lceil\lg m\rceil)\sum_{i=1}^{m}{n_i d_{i}^{2}} \, .
\end{displaymath}
The cost of line~\ref{one:more:bit} is very modest because we use \emph{truncations} rather than general approximations in line~\ref{use:truncations} for the leader to compute $p_x(t_0)$ for all \mbox{$x\in\Dom$}.
Indeed, it suffices to obtain a single additional bit of precision in the real and imaginary parts of each entry defining matrix $M_{i\subX_i}$ from each custodian~\mbox{$i\in[m]$}.
The cost of this update is simply
\begin{equation}\label{eq:simple:gamma}
\gamma=m + \sum_{i=1}^{m}{d_{i}^{2}}
\end{equation}
bits of communication, where the addition of $m$ is to account for the leader needing to request new bits from the custodians. This is a nice example of what we meant by
``it~could be that bits previously communicated can be reused'' in line~\ref{we:can:reuse} of Algorithm~\ref{algo:vnadaptedone}.

Putting it all together in Eq.~(\ref{eq:tradeoff}),
the total expected number of bits communicated in Algorithm~\ref{algo:simulation} in order to sample exactly according to the quantum probability distribution is
\begin{align}
\mathbf{E}(Z) & \leq \delta(t_0) + 3 \gamma \big( 1+2^{1-t_0} n \big) \nonumber \\
& \leq (t_0+2+\lceil2\lg d\rceil+2 \lceil\lg m\rceil)\sum_{i=1}^{m}{n_i d_{i}^{2}} + 3 \big( 1+2^{1-t_0} n \big) \bigg( m+ \sum_{i=1}^{m}{d_{i}^{2}} \bigg) \, . \label{eq:final_simul}
\end{align}

We are finally in a position to choose the value of parameter~$t_0$.
First note that \mbox{$n = \prod_{i=1}^m n_i \geq 2^m$}.
Therefore, any constant choice of $t_0$ will entail an expected amount of communication that is exponential in~$m$ because of the right-hand term in Eq.~(\ref{eq:final_simul}).
At~the other extreme, choosing \mbox{$t_0=n$} would also entail an expected amount of communication that is exponential in~$m$, this time because of the left-hand term in Eq.~(\ref{eq:final_simul}).
A~good compromise is to choose \mbox{$t_0 = \lceil \, \lg n \rceil$}, which results in \mbox{$1 \le C \le 3$} according to Eq.~(\ref{equa:uppbndC}), because in that case \mbox{$2^{t_0} \geq n$} and therefore
\begin{displaymath}
1 \leq C \le 1+2^{1-t_0} n = 1 + \frac{2n}{2^{t_0}} \le 3 \, ,
\end{displaymath}
so that Eq.~(\ref{eq:final_simul}) becomes
\begin{equation}
\mathbf{E}(Z) \leq (\lceil \, \lg n \rceil+\lceil2\lg d\rceil+2 \lceil\lg m\rceil+2)\sum_{i=1}^{m}{n_i d_{i}^{2}} + 9 \bigg( m+ \sum_{i=1}^{m}{d_{i}^{2}} \bigg) \, . \label{eq:final_simul_tradeoff}
\end{equation}
In case all the $n_i$'s and $d_i$'s are upper-bounded by some constant~$\xi$, we have that
\mbox{$n = \prod_{i=1}^{m}{n_i} \leq \xi^m$}, hence \mbox{$\lg n \leq m \lg \xi$},
similarly \mbox{$\lg d \leq m \lg \xi$},
and also \mbox{$\sum_{i=1}^{m}{n_i d_{i}^{2}} \leq m \xi^3$}.
It~follows that
\begin{equation}\label{eq:bounded}
\mathbf{E}(Z) \leq (3 \xi^3 \lg \xi) m^2 + O(m \log m) \, ,
\end{equation}
which is in the order of~$m^2$, thus matching with our most general method the result that was already known for the very specific case of simulating the quantum \mbox{$m$-par}\-tite GHZ distribution~\cite{BrDeGr2015}.

\section{Practical Implementation Using a Source of Discrete Randomness}\label{app:bitmodel}

In practice, we cannot work with continuous random variables since our computers have finite storage capacities and finite precision arithmetic. Furthermore, the generation of uniform continuous random variables does not make sense computationally speaking and we must adapt Algorithms~\ref{algo:vnadaptedone} and~\ref{algo:simulation} to work in a finite world.

For this purpose, recall that $U$ is a uniform continuous random variable on \mbox{$[0,1]$} used in all the algorithms seen so~far.
For~each \mbox{$i \geq 1$}, let~$U_i$ denote the $i^{\text{th}}$ bit in the binary expansion of $U$, so that
\begin{displaymath}
U=0.U_{1}U_{2}\cdots=\sum_{i=1}^{\infty}{U_{i}2^{-i}}.
\end{displaymath}
We acknowledge the fact that the $U_i$'s are not uniquely defined in case
\mbox{$U=j/2^{k}$} for integers \mbox{$k>0$} and \mbox{$0 < j < 2^k$}, but we only mention this phenomenon to ignore it since it occurs with probability~$0$ when $U$ is uniformly distributed on~\mbox{$[0,1]$}.
We denote the \mbox{$t$-bit} truncation of~$U$ by $U[t]$:
\begin{displaymath}
U[t]\isdefdisp \lfloor 2^t U \rfloor / 2^t = \sum_{i=1}^{t}{U_{i} 2^{-i}} \, .
\end{displaymath}
For all \mbox{$t \geq 1$}, we have that
\begin{equation}\label{eq:trunc:U}
U[t]\leq U < U[t]+2^{-t}.
\end{equation}

We modify Algorithm~\ref{algo:vnadaptedone} into Algorithm~\ref{algo:vnadaptedone_bitmodel} as follows, leaving to the reader the corresponding
modification of Algorithm~\ref{algo:simulation}, thus yielding a practical protocol for the simulation of general entanglement under arbitrary measurements.

\begin{algorithm}
\caption{Modified rejection algorithm with discrete randomness source -- Protocol for the leader}\label{algo:vnadaptedone_bitmodel}
{\fontsize{10}{18}\selectfont
\begin{algorithmic}[1]
\REQUIRE Value of $t_0$
\STATE Compute $p_x(t_0)$ for each $x \in \Dom$ \\[-1ex]
\COMMENT{The leader needs information from the custodians in order to compute these approximations}
\STATE Compute $C$ and $q=(q_x)_{x\in\Dom}$ as per Eqs.~(\ref{equa:C}) and~(\ref{equa:qiko})
\STATE Sample $X$ according to $q$\label{gen:X:discrete}
\STATE $U[0]\leftarrow 0$
\FOR{$t=1$ \TO $t_0-1$} \label{initial:for:loop:discrete}
\STATE Generate i.i.d.~unbiased bit $U_t$\label{gen:Ubit1:discrete}
\STATE $U[t] \leftarrow U[t-1] + U_t \, 2^{-t}$
\ENDFOR
\FOR{$t=t_0$ \TO $\infty$\label{for:loop:discrete}}
\STATE Generate i.i.d.~unbiased bit $U_t$\label{gen:Ubit2:discrete}
\STATE $U[t] \leftarrow U[t-1] + U_t \, 2^{-t}$
\IF{$\big(U[t]+2^{-t}\big)Cq_{\subX}\leq p_{\subX}(t)-2^{-t}$}\label{test:to:accept:discrete}
\RETURN $X$\label{line:accept:discrete} \COMMENT{$X$ is accepted}
\ELSIF{$U[t] Cq_{\subX} > p_{\subX}(t)+2^{-t}$}\label{test:to:reject:discrete}
\STATE \textbf{go to}~\ref{gen:X:discrete}\label{line:reject:discrete} \COMMENT{$X$ is rejected}
\ELSE
\STATE Continue the \textbf{for} loop \\[-1ex]
\COMMENT{We cannot decide to accept or reject because \mbox{$-(1+Cq_{\subX}) 2^{-t} < U[t] C q_{\subX} - p_{\subX}(t) \leq 2^{-t}$}\,; \\[-1ex]
communication may be required in order for the leader to compute \mbox{$p_{\subX}(t+1)$}\,; \\[-1ex]
it~could be that bits previously communicated to compute $p_{x}(t)$ can be reused.}\label{we:can:reuse:discrete}
\ENDIF
\ENDFOR
\end{algorithmic}}
\end{algorithm}

\begin{Theorem}\label{thm:discrete}
Algorithm~\ref{algo:vnadaptedone_bitmodel} is correct, i.e., it terminates and returns \mbox{$X=x$} with probability~$p_x$. Furthermore, let $T$ be the random variable that denotes the value of variable $t$ upon termination of any instance of the \textup{\textbf{for}} loop that starts at line~\ref{for:loop:discrete}, whether it terminates in rejection or acceptation. Then
\begin{displaymath}
\mathbf{E}(T)\leq t_0+3+2^{-t_0} \, .
\end{displaymath}
\end{Theorem}

\begin{proof}
This is very similar to the proof of Theorem~\ref{thm:main}, so let us concentrate on the differences.
First note that it follows from Eq.~(\ref{eq:trunc:U}) and the fact that \mbox{$ | p_{\subX}(t)-p_{\subX} | \le 2^{-t} $} that
\begin{displaymath}
\big(U[t]+2^{-t}\big)Cq_{\subX}\leq p_{\subX}(t)-2^{-t}
\implies UCq_{\subX} \leq p_{\subX}(t)-2^{-t}
\implies UCq_{\subX} \leq p_{\subX}
\end{displaymath}
and
\begin{displaymath}
U[t] Cq_{\subX} > p_{\subX}(t)+2^{-t}
\implies UCq_{\subX} > p_{\subX}(t)+2^{-t}
\implies UCq_{\subX} > p_{\subX} \, .
\end{displaymath}
Therefore, whenever $X$ is accepted at line~\ref{line:accept:discrete}
(resp.~rejected at line~\ref{line:reject:discrete}), it would also have been accepted
(resp.~rejected) in the original von Neumann algorithm, which shows sampling correctness.
Conversely, whenever we reach a value of \mbox{$t \geq t_0$} such that
\mbox{$\big(U[t]+2^{-t}\big)Cq_{\subX} > p_{\subX}(t)-2^{-t}$} and \mbox{$U[t] Cq_{\subX} \leq p_{\subX}(t)+2^{-t}$}, we do not have enough information to decide whether to accept or reject, and therefore we reach line~\ref{we:can:reuse:discrete}, causing $t$ to increase.
This happens precisely when
\begin{displaymath}
-(1+Cq_{\subX}) 2^{-t} < U[t] C q_{\subX} - p_{\subX}(t) \leq 2^{-t} \, .
\end{displaymath}
To obtain an upper bound on $\mathbf{E}(T)$, we mimic the proof of Theorem~\ref{thm:main}, but in the discrete rather than continuous regime.
In~particular, for any \mbox{$x\in\Dom$} and \mbox{$t \geq t_0$},
\begin{align}
\mathbf{P}\!\left\{T>t~\rvert~X=x\right\}
& \leq \mathbf{P}\!\left\{-(1+Cq_{x}) 2^{-t} < U[t] C q_{x} - p_{x}(t) \leq 2^{-t}\right\} \nonumber \\
&= \mathbf{P}\!\left\{ p_{x}(t)-(1+Cq_{x}) 2^{-t} < U[t] C q_{x} \leq p_{x}(t)+2^{-t} \right\} \nonumber \\
&= \mathbf{P}\!\left\{ \frac{2^t p_{x}(t)}{C q_{x}}-\frac{1+Cq_{x}}{C q_{x}} < 2^t U[t] \leq \frac{2^t p_{x}(t)}{C q_{x}}+\frac{1}{C q_{x}} \right\} \nonumber \\
&\le \left[ \left( \frac{2^t p_{x}(t)}{C q_{x}}+\frac{1}{C q_{x}} \right) - \left( \frac{2^t p_{x}(t)}{C q_{x}}-\frac{1+Cq_{x}}{C q_{x}} \right) + 1\right] 2^{-t} \label{eq:nearly:there} \\
&= 2 \left( 1+ \frac{1}{Cq_x} \right) 2^{-t}
\leq 2^{t_0-t+1} + 2^{1-t}~~~~~~\left(\text{because}~ Cq_x \geq 2^{-t_0}\right)\, . \label{prob_nohalt_discrete}
\end{align}
To understand Eq.~(\ref{eq:nearly:there}), think of $2^t U[t]$ as an integer chosen randomly and uniformly between $0$ and \mbox{$2^t-1$}. The probability that it falls within some real interval \mbox{$(a,b]$} for \mbox{$a<b$} is equal to $2^{-t}$ times the number of integers between $0$ and \mbox{$2^t-1$} in that interval, the latter being upper-bounded by the number of unrestricted integers in that interval, which is at most \mbox{$b-a+1$}.

Noting how similar Eq.~(\ref{prob_nohalt_discrete}) is to the corresponding Eq.~(\ref{prob_nohalt}) in the analysis of Algorithm~\ref{algo:vnadaptedone}, it~is not surprising that the expected value of~$T$ will be similar as well. And~indeed, continuing as in the proof of Theorem~\ref{thm:main},
without belabouring the details,
\begin{align}
\mathbf{E}(T~\rvert~X=x)&=\sum_{t=0}^{\infty}{\mathbf{P}\{T>t~\rvert~ X=x\}} \nonumber \\
&=\sum_{t=0}^{t_0+1}{\mathbf{P}\{T>t~\rvert~ X=x\}} + \!\!\!\!\sum_{t=t_0+2}^{\infty}\!\!{\mathbf{P}\{T>t~\rvert~ X=x\}} \nonumber \\
&\leq t_0 + 2 + 2^{t_0+1}\!\!\!\!\sum_{t=t_{0}+2}^{\infty}\!\!{2^{-t}} + 2\!\!\!\!\sum_{t=t_{0}+2}^{\infty}\!\!{2^{-t}} = t_0+3+2^{-t_0} \label{eq:expT:discrete} \, .
\end{align}
We~conclude that \mbox{$\mathbf{E}(T) \leq t_0+3+2^{-t_0}$} without condition since Eq.~(\ref{eq:expT:discrete}) does not depend on~$x$.
\end{proof}

The similarity between Theorems~\ref{thm:main} and~\ref{thm:discrete} means that there is no significant additional cost in the amount of communication required to achieve remote sampling in the random bit model. i.e., if~we consider a realistic scenario in which the only source of randomness comes from i.i.d.~unbiased bits, compared to an unrealistic scenario in which continuous random variables can be drawn. For~instance, the reasoning that led to Eq.~(\ref{eq:tradeoff}) applies \emph{mutatis mutandis} to conclude that the expected number $Z$ of bits that need to be communicated to achieve remote sampling in the random bit model is
\begin{displaymath}
\mathbf{E}(Z) \leq \delta(t_0) + \big( 3+2^{-t_0} \big) \big( 1+2^{1-t_0} n \big) \gamma \, ,
\end{displaymath}
where $\delta$ and $\gamma$ have the same meaning as in Section~\ref{sect:modifVN}.

If we use the random bit approach for the general simulation of quantum entanglement studied in Section~\ref{sect:simquantent}, choosing \mbox{$t_0 = \lceil \, \lg n \rceil$} again, Eq.~(\ref{eq:final_simul_tradeoff}) becomes
\begin{equation}\label{eq:final_simul_tradeoff:discrete}
\mathbf{E}(Z) \leq (\lceil \, \lg n \rceil+\lceil2\lg d\rceil+2 \lceil\lg m\rceil+2)\sum_{i=1}^{m}{n_i d_{i}^{2}} + 3 (3+1/n) \bigg( m+ \sum_{i=1}^{m}{d_{i}^{2}} \bigg) \, ,
\end{equation}
which reduces to the identical
\begin{displaymath}
\mathbf{E}(Z) \leq (3 \xi^3 \lg \xi) m^2 + O(m \log m)
\end{displaymath}
in case all the $n_i$'s and $d_i$'s are upper-bounded by some constant~$\xi$,
which again is in the order of~$m^2$.

In addition to communication complexity, another natural efficiency measure in the random bit model concerns the \emph{expected number of random bits} that needs to be drawn in order to achieve sampling. Randomness is needed in lines \ref{gen:X:discrete}, \ref{gen:Ubit1:discrete} and~\ref{gen:Ubit2:discrete} of Algorithm~\ref{algo:vnadaptedone_bitmodel}.
A~single random bit is required each time lines \ref{gen:Ubit1:discrete} and~\ref{gen:Ubit2:discrete} are entered, but line~\ref{gen:X:discrete} calls for sampling $X$ according to distribution~$q$.
Let~$V_i$ be the random variable that represents the number of random bits needed on the $i^{\text{th}}$ passage through line~\ref{gen:X:discrete}.
For~this purpose, we use the algorithm introduced by Donald Knuth and Andrew Chi-Chih Yao~\cite{KnuthYao1976}, which enables sampling within any finite discrete probability distribution in the random bit model by using an expectation of no more than two random bits in addition to the Shannon binary entropy of the distribution. Since each such sampling is independent from the others, it follows that $V_i$ is independently and identically distributed as a random variable $V$ such that
\begin{equation}\label{eq:KY}
\mathbf{E}(V) \leq 2+H(q) \leq 2+\lg n \, ,
\end{equation}
where $H(q)$ denotes the binary entropy of~$q$, which is never more than the \mbox{base-two} logarithm of the number of atoms in the distribution, here~$n$.

Let~$R$ be the random variable that represents the number of random bits drawn when running Algorithm~\ref{algo:vnadaptedone_bitmodel}.
Reusing the notation of Section~\ref{sect:modifVN},
let $S$ be the random variable that represents the number of times variable $X$ is sampled at line~\ref{gen:X:discrete} and let $T_i$ be the random variable that represents the value of variable $T$ upon termination of the $i^{\text{th}}$ instance of the \textbf{for} loop starting at line~\ref{for:loop:discrete}, for $i\in\{1,\ldots,S\}$.
The random variables $T_i$ are independently and identically distributed as the random variable $T$ in Theorem~\ref{thm:discrete} and the expected value of $S$ is~$C$.
Since one new random bit is generated precisely each time variable $t$ is increased by~$1$ in any pass through either \textbf{for} loops (line~\ref{initial:for:loop:discrete} or~\ref{for:loop:discrete}), we simply have
\begin{displaymath}
R = \sum_{i=1}^{S}{\left( V_i+T_i \right)} \, .
\end{displaymath}
By~virtue of Eqs.~(\ref{equa:uppbndC}) and~(\ref{eq:KY}), Theorem~\ref{thm:discrete}, and using Wald's identity again, we conclude:
\begin{align*}
\mathbf{E}(R) &= \mathbf{E}(S) \left( \mathbf{E}(V) + \mathbf{E}(T) \right) \\
&\leq \big(1+2^{1-t_0} n \big) \big( \lg n + t_0+5+2^{-t_0} \big) \, .
\end{align*}

Taking \mbox{$t_0=\lceil\,\lg n \rceil$} again, remote sampling can be completed using an expected number of random bit in $O(\lg n)$, with a hidden multiplicative constant no larger than~$6$.
The hidden constant can be reduced arbitrarily close to~$2$ by taking \mbox{$t_0=\lceil\,\lg n \rceil+a$} for an arbitrarily large constant~$a$.
Whenever target distribution $p$ has close to full entropy, this is only twice the optimal number of random bits that would be required according to the Knuth-Yao lower bound~\cite{KnuthYao1976} in the usual case when full knowledge of $p$ is available in a central place rather than having to perform remote sampling. 
Note however that if our primary consideration is to optimize communication for the classical simulation of entanglement, as~in Section~\ref{sect:simquantent}, choosing \mbox{$t_0=\lceil\,\lg n \rceil-a$} would be a better idea because 
the summation in the left-hand term of Eq.~(\ref{eq:final_simul}) dominates that of the right-hand term.
For this inconsequential optimization, $a$ does not have to be a constant but it should not exceed~$\lg (\xi m)$, where $\xi$ is our usual upper bound on the number of possible outcomes for each participant (if~it exists), lest the right-hand term of Eq.~(\ref{eq:final_simul}) overtake the left-hand term.
Provided $\xi$ exists, the expected number of random bits that need to be drawn is linear in the number of participants.

The need for continuous random variables was not the only unrealistic assumption in Algorithms \ref{algo:vonNeumannOriginal} to~\ref{algo:simulation}. We~had also assumed implicitly that custodians know their private parameters precisely (and that the leader knows exactly each entry of density matrix~$\rho$ in Section~\ref{sect:simquantent}). This could be reasonable in some situations, but it could also be that some of those parameters are transcendental numbers or the result of applying transcendental functions to other parameters, for example~$\cos \pi/8$. More interestingly, it could be that the actual parameters are spoon-fed to the custodians by \emph{examiners}, who want to test the custodians' ability to respond appropriately to unpredictable inputs. However, all we need is for the custodians to be able to obtain their own parameters with arbitrary precision, so that they can provide that information to the leader upon request. For~example, if a parameter is $\pi/4$ and the leader requests a \mbox{$k$-bit} approximation of that parameter, the custodian can compute some number $\hat{x}$ such that \mbox{$|\hat{x}-\pi/4| \leq 2^{-k}$} and provide it to the leader. For~communication efficiency purposes, it is best if $\hat{x}$ itself requires only $k$ bits to be communicated,
or perhaps one more (for the sign) in case the parameter is constrained to be between $-1$ and~$1$. It~is even better if the custodian can supply a \mbox{$k$-bit} \emph{truncation} because this enables the possibility to upgrade it to a \mbox{$(k+1)$-bit} truncation by the transmission of a single bit upon request from the leader, as we have done explicitly for the simulation of entanglement at line~\ref{one:more:bit} of Algorithm~\ref{algo:simulation} in Section~\ref{sect:simquantent}.

Nevertheless, it may be impossible for the custodians to compute truncations of their own parameters in some cases, even when they can compute arbitrarily precise approximations.
Consider for instance a situation in which one parameter held by a custodian is \mbox{$x=\cos\theta$} for some angle $\theta$ for which he can only obtain arbitrarily precise truncations. Unbeknownst to the custodian, \mbox{$\theta=\pi/3$} and therefore~\mbox{$x=1/2$}.  No~matter how many bits the custodian obtains in the truncation of $\theta$, however, he can never decide whether \mbox{$\theta<\pi/3$} or \mbox{$\theta\ge\pi/3$}. In~the first case, \mbox{$x<1/2$} and therefore the \mbox{$1$-bit} truncation of $x$ should be~$0$, whereas in the second (correct) case, \mbox{$x\ge1/2$} and therefore the \mbox{$1$-bit} truncation of $x$ is~$1/2$ (or~$0.1$~in binary). Thus, the custodian will be unable to respond if the leader asks him for a \mbox{$1$-bit} truncation of~$x$, no matter how much time he spends on the task. In~this example, by contrast, the custodian can supply the leader with arbitrarily precise \emph{approximations} of~$x$ from appropriate approximations of~$\theta$. Should a situation like this occur, for instance in the simulation of entanglement, there would be two solutions. The first one is for the custodian to transmit increasingly precise truncations of~$\theta$ to the leader and let \emph{him} compute the cosine on~it. This approach is only valid if it is known at the outset that the custodian's parameter will be of that form, which was essentially the solution taken in our earlier work on the simulation of the quantum \mbox{$m$-par}\-tite GHZ distribution~\cite{BrDeGr2015}. The more general solution is to modify the protocol and declare that custodians can send arbitrary approximations to the leader rather than truncations. The consequence on Algorithm~\ref{algo:simulation} is that line~\ref{one:more:bit} would become much more expensive since each custodian $i$ would have to transmit a fresh \mbox{one-bit}-better approximation for the real and imaginary parts of the $d_i^2$ entries defining matrix~$M_{i\subX_i}$. As~a result, efficiency parameter $\gamma(t)$ in Eq.~(\ref{eq:if:no:trunc}) would become
\begin{displaymath}
\gamma(t) = m + (t+2+\lceil2\lg d\rceil+2 \lceil\lg m\rceil)\sum_{i=1}^{m}{d_{i}^{2}} \, ,
\end{displaymath}
which should be compared with the much smaller (constant) value of $\gamma$ given in Eq.~(\ref{eq:simple:gamma}) when truncations of the parameters are available.
Nevertheless, taking \mbox{$t_0 = \lceil \, \lg n \rceil$} again, this increase in $\gamma(t)$ would make no significant difference in the total number of bits transmitted for the simulation of entanglement because it would increase only the right-hand term in Eqs.~(\ref{eq:final_simul_tradeoff}) and~(\ref{eq:final_simul_tradeoff:discrete}), but not enough to make it dominate the left-hand term.
All counted, we still have an expected number of bits transmitted that is upper-bounded by
\mbox{$(3 \xi^3 \lg \xi) m^2 + O(m \log m)$}
whenever all the $n_i$'s and $d_i$'s are upper-bounded by some constant~$\xi$,
which again is in the order of~$m^2$.

\section{Conclusion, Discussion and Open Problems}\label{sc:conclusion}

We have introduced and studied the general problem of sampling a discrete probability distribution characterized by parameters that are scattered in remote locations. Our main goal was to minimize the amount of communication required to solve this conundrum. We~considered both the unrealistic model in which arithmetic can be carried out with infinite precision and continuous random variables can be sampled exactly, and the more reasonable \emph{random bit model} studied by Knuth and Yao, in which all arithmetic is carried out with finite precision and the only source of randomness comes from independent tosses of a fair coin. For~a small increase in the amount of communication, we~can fine-tune our technique to require only twice the number of random bits that would be provably required in the standard context in which all the parameters defining the probability distribution would be available in a single location, provided the entropy of the distribution is close to maximal.

When our framework is applied to the problem of simulating quantum entanglement with classical communication in its most general form, we find that an expected number of $O(m^2)$ bits of communication suffices when there are $m$ participants and each one of them (in~the simulated world) is~given an arbitrary quantum system of bounded dimension and asked to perform an arbitrary generalized measurement (\POVM) with a bounded number of possible outcomes. This result generalizes and supersedes the best approach previously known in the context of multi-party entanglement, which was for the simulation of the \mbox{$m$-par}\-tite GHZ state under projective measurements~\cite{BrDeGr2015}. Our technique also applies without the boundedness condition on the dimension of individual systems and the number of possible outcomes per party, provided those parameters remain finite.

Our work leaves several important questions open.
Recall that our approach provides a bounded amount on the \emph{expected} communication required to perform exact remote sampling.
The most challenging open question is to determine if it is possible to achieve the same goal with a guaranteed bounded amount of communication \emph{in~the worst case}. If~possible, this would certainly require the participants to share ahead of time the realization of random variables, possibly even continuous~ones.
Furthermore, a radically different approach would be needed since we had based ours on the von Neumann rejection algorithm,
which has intrinsically no worst-case upper bound on its performance.
This task may seem hopeless, but it has been shown to be possible for special cases of entanglement simulation in which the remote parameters are taken from a continuum of possibilities~\cite{bct99,TB03}, despite earlier ``proofs'' that it is impossible~\cite{maudlin92}.

A much easier task would be to consider other communication models, in which communication is no longer restricted to being between a single leader and various custodians. Would there be an advantage in communicating through the edges of a complete graph? Obviously, the maximum possible saving in terms of communication would be a factor of~2 since any time one participant wants to send a bit to some other participant, he can do so via the leader. However, if we care not only about the total number of bits communicated, but also the \emph{time} it takes to complete the protocol in a realistic model in which each party is limited to sending and receiving a fixed number of bits at any given time step, parallelizing communication could become valuable. We~had already shown in Ref.~\cite{BrDeGr2015} that a parallel model of communication can dramatically improve the time needed to sample the \mbox{$m$-par}\-tite GHZ distribution.
Can this approach be generalized to arbitrary remote sampling settings?

Finally, we would like to see applications for remote sampling outside the realm of quantum information science.


\vspace{6pt} 

\funding{The work of G.B.\ is supported in part by the Canadian Institute for Advanced Research, the Canada Research Chair program, Canada's Natural Sciences and Engineering Research Council (\textsc{nserc}) and Qu\'ebec's Institut transdisciplinaire d'information quantique.
The work of L.D.\ is supported in part by \textsc{nserc}.}

\acknowledgments{The authors are very grateful to Nicolas Gisin for his interest in this work and the many discussions we have had with him on this topic in the past decade.
Marc Kaplan has also provided important insights in earlier work on the simulation of entanglement.}

\abbreviations{The following abbreviations are used in this manuscript:\\
\noindent 
\begin{tabular}{@{}ll}
i.i.d. & independent identically distributed \\
GHZ & Greenberger-Horne-Zeilinger \\
POVM & positive-operator valued measure
\end{tabular}}

\reftitle{References}

\end{document}